# Designing a family of 2D kagome monolayer $B_{18}S_8$, $B_{18}S_8H_2$, $B_{18}S_6X_2$ (X=Cl, Br, I) with tunable Dirac cones and high Fermi velocity

Su-Yang Shen,[1] En-Qi Bao,[1] Xing-Yu Wang,[1,2] Jiafu Wang,[1,2,*] and Jun-Hui Yuan,[1,*]

[1]School of Physics and Mechanics, Wuhan University of Technology, Wuhan 430070, China

[2]School of Materials and Microelectronics, Wuhan University of Technology, Wuhan 430070, China

**[*]Corresponding Author**

E-mail: jasper@whut.edu.cn (J. Wang); yuanjh90@163.com (J.-H. Yuan)

**ABSTRACT**

Two-dimensional (2D) kagome materials have become a hot research topic in the current scientific community due to their unique electronic structural properties, and the design of novel 2D kagome materials represents a significant exploration direction in this field. In this study, by employing the “1+3” design strategy, surface passivation and charge balance strategies, we successfully designed a novel family pf 2D kagome material $B_{18}S_8$, $B_{18}S_8H_2$ and $B_{18}S_6X_2$ (X = Cl, Br, I). Electronic structure analysis revealed that although $B_{18}S_8$ exhibits excellent kagome band characteristics, its Dirac cone is located approximately 1 eV above the Fermi level, making it difficult to utilize. However, by surface hydrogen passivation, the Dirac cone can be effectively adjusted to the Fermi level. Further research found that introducing halogen atoms to replace surface sulfur atoms can similarly adjust the position of the Dirac cone to the Fermi level. The Fermi velocities near the Dirac cone for these five materials reach as high as 2.69 to $3.07\times10^5$ m/s. Additionally, spin-orbit coupling can open a bandgap of approximately 20 to 55 meV at the Dirac cone. Our research not only provides an outstanding example for the design of 2D boron-based kagome materials but also fully demonstrates the immense potential of such materials in the electronics field.

## 1. Introduction

The rise of two-dimensional (2D) materials has shattered the performance boundaries of traditional three-dimensional (3D) materials. With their atomic-level thickness, these materials exhibit significant quantum confinement effects, high specific surface areas, and tunable interfacial properties, serving as a crucial bridge connecting fundamental physics research with high-tech applications[1]. Since the groundbreaking exfoliation of graphene[2], the family of 2D materials has continuously expanded[3]. Leveraging the unique valence electron configuration and versatile bonding modes of boron, boron-based 2D materials have formed a diverse system encompassing elemental forms, compounds, and composites[4–6]. These materials demonstrate physicochemical properties that are difficult to replace by typical 2D materials such as graphene, gradually emerging as a research frontier in materials science, condensed matter physics, and device engineering[7–9].

Boron atoms exhibit flexible bonding capabilities, forming both $sp^2$-hybridized planar coordination and 2D-derived 3D topological network structures through multicenter bonds[10]. This flexibility endows boron-based 2D materials with structural diversity and highly tunable electronic, optical, mechanical, and catalytic properties[11–14]. For instance, borophene subverts the traditional 3D bonding mode of elemental boron, featuring ordered vacancies in its monolayer that form a unique “hole” topology[15,16]. This structure grants it high mechanical strength and carrier mobility, enables the modulation of electronic states, and holds immense potential in high-speed flexible electronics and other fields[17–19]. Recently, the research scope of boron-based materials has been successfully extended to an emerging hot field—kagome materials—a breakthrough that has significantly stimulated research enthusiasm among scientists[20–23]. For example, in the study of double-kagome-layered metallic borophene, researchers have discovered higher-order van Hove singularities and remarkable superconducting properties[24]. More notably, through surface passivation, these materials can be further tuned into semiconductors $B_3X$ (X = O, S, Se, and Te) with ultra-high stability, while preserving typical kagome material features such as flat bands and Dirac cones in their electronic structures[25,26].

In this study, we utilized the highly stable 2D boride $B_3S$ proposed in our previous work as the parent material and employed an innovative "1+3" design strategy to thoroughly explore theoretical design pathways for novel 2D kagome materials[27–29]. Subsequently, based on first-principles calculations, we systematically investigated the stability, electronic structure, and other critical properties of the designed monolayers. The results demonstrate that through surface passivation or element substitution, the Dirac cone can be effectively and precisely tuned to the Fermi level, leading to the acquisition of semimetallic materials with high Fermi velocities compared to that of graphene. This discovery provides a highly valuable paradigm for the research of novel boron-based kagome materials.

## 2. Design Strategy

We first provide a detailed explanation of the material design strategy employed in this work. **Figure 1a** illustrates the crystal structure of the 2×2×1 supercell of $B_{18}S_6$. $B_{18}S_6$

features a typical hexagonal structure, composed of atomic clusters "$B_6$" and sulfur (S) atoms. From a chemical formula perspective, it can also be represented as $(B_6)_1S_2$, and its crystal structure is consistent with that of the *T*-phase molybdenum disulfide ($MoS_2$). Based on the "1+3" design strategy, within the $B_{18}S_6$ structure, if the "$B_6$" atomic clusters at the lattice vertices are removed, the remaining three "$B_6$" atomic clusters will spontaneously form a typical kagome lattice, as shown in **Figure 1b**. The chemical formula of the newly generated kagome material is $B_{18}S_8$, in which S atoms occupy two different non-equivalent sites: one is the three-coordinate S1 atoms located at the center of the "$B_6$" kagome lattice (as indicated by the pink dashed box in **Figure 1b**); the other is the two-coordinate S2 atoms situated on the outer side of the kagome lattice.

In $B_{18}S_6$, the "$B_6$" atomic clusters exist as cations. After removing the "$B_6$" atomic clusters, the resulting $B_{18}S_8$ may have an "excess" of electrons in the system due to the presence of cation vacancies. In view of this, without altering the symmetry of the kagome lattice, we performed surface passivation treatment on $B_{18}S_8$, that is, using hydrogen (H) atoms to passivate the three-coordinate S2 atoms. This operation can effectively reduce the number of electrons in the system, leading to the formation of $B_{18}S_8H_2$, as shown in **Figure 1c**. Through this approach, we further achieved tuning of the electronic structure of the system. Subsequent electronic structure calculations strongly confirm the feasibility and effectiveness of this design strategy, with details provided in Section 3. From a charge perspective, after hydrogen passivation, $B_{18}S_8H_2$ forms "-SH" groups with a "-1" valence. Halogen atoms inherently possess a natural -1 valence. Based on this principle of charge balance, we further replaced the "-SH" groups with halogen elements X (X = Cl, Br, I), successfully obtaining $B_{18}S_6X_2$, as shown in **Figure 1d**. It should be noted that $B_{18}S_6X_2$ can also be obtained by element substitution of the three-coordinate S1 atoms in $B_{18}S_8$.

Based on the aforementioned material design strategy, we successfully predicted a total of five novel 2D boron-based kagome materials, namely $B_{18}S_8$, $B_{18}S_8H_2$, and $B_{18}S_2X_2$ (X=Cl, Br, I). Subsequently, we conducted systematic and comprehensive computational analyses of the stability, electronic structure, and other key properties of these material systems using the VASP software[30,31]. Specific computational methods and parameter settings can be found in **Note 1** of the **Supplementary Material**.

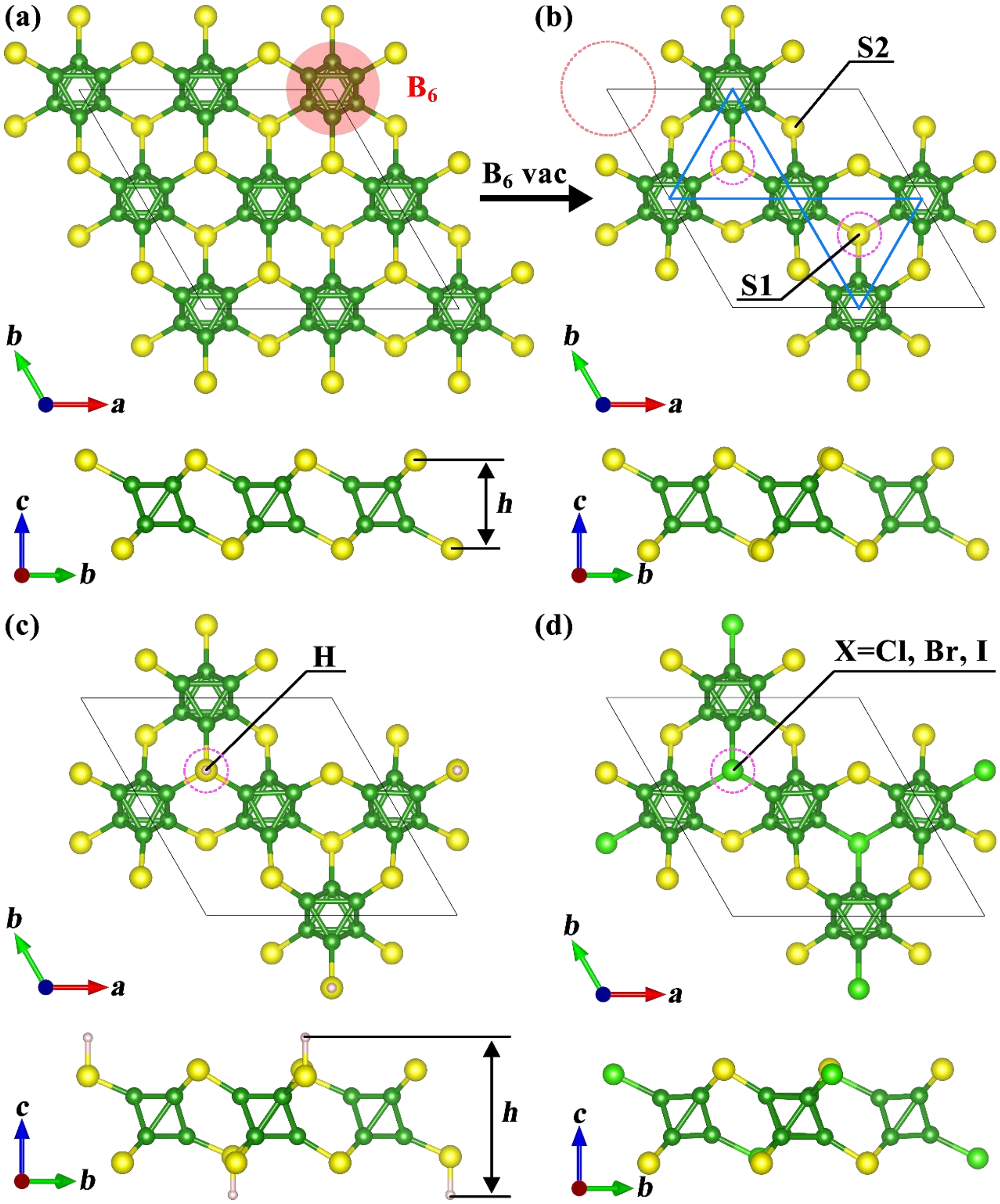

**Figure 1.** Schematic illustrations of the design concept for boron-based 2D kagome materials. (a) Top and (b) side view of the $B_{18}S_6$ monolayer, showing the basic units of $B_6$ clusters (pink circle). (b) Top and side view of the $B_{18}S_8$ monolayer with marked the kagome lattice and $B_6$ vacancy. Top and side view of the (c) $B_{18}S_8H_2$ and (d) $B_{18}S_6X_2$ monolayer. The atomic thickness $h$ is marked in the figures.

### 3. Results and discussions

Based on the design strategies, five novel boron-based kagome materials can theoretically be constructed. The lattice constants *a*/*b*, atomic layer thicknesses *h*, bond and lengths *l* obtained after structural optimization are listed in **Table 1**. After introducing a "$B_6$" vacancy into $B_{18}S_6$, its lattice constant decreases from 9.14 Å to 9.056 Å; however, the atomic layer thickness shows a slight increasing trend, rising from 3.190 Å to 3.227 Å. Meanwhile, due to the transformation of the originally 3-coordinated S atoms in $B_{18}S_6$ into partially 3-coordinated S1 atoms and 2-coordinated S2 atoms, this transformation leads to a shortening of the B-S2 bond length from 1.88 Å to 1.809 Å, with a reduction of approximately 3.77%, while the B-S1 bond length remains essentially unchanged. Another relatively significant change is observed in the B-B bonds. In $B_{18}S_8$, the bond lengths of both the in-layer and inter-layer B-B bonds increase, with the maximum increases for the in-layer and inter-layer B-B bonds being approximately 3.71% and 1.68%, respectively.

After hydrogen passivation treatment, the lattice constant of the material increases significantly from 9.056 Å to 9.449 Å; while the bond lengths of the in-layer and inter-layer B-B bonds change very little, the B-S1 bond length decreases. When the S1 atoms are replaced with halogen atoms, except for a slight increase in the lattice constant of $B_{18}S_6Br_2$ (by 2.14%), the lattice constants of $B_{18}S_6Cl_2$ and $B_{18}S_6I_2$ both decrease slightly compared to $B_{18}S_8$. However, it is noteworthy that the atomic layer thicknesses of these materials are all larger than that of $B_{18}S_8$ and exhibit a periodic increase with different X atoms. In $B_{18}S_6X_2$, the bond lengths of the in-layer and inter-layer B-B bonds, as well as the B-S2 bond length, remain essentially consistent, while the B-X bond length shows a periodic increase with different X atoms. Overall, $B_{18}S_8H_2$ and $B_{18}S_6X_2$ derived from $B_{18}S_8$ maintain good structural consistency. This excellent structural consistency undoubtedly provides extremely favorable conditions for potential conversion between different material systems.

However, the experimental feasibility of such materials remains the core prerequisite and the foremost issue to be addressed for their transition from theoretical design to practical synthesis. To this end, we systematically evaluated the kinetic, thermal, and mechanical stabilities of the aforementioned five kagome materials through phonon dispersion spectra, *ab initio* molecular dynamics (AIMD) simulations, and elastic constant calculations. As shown in **Figures 2a-2c** and **Figure S1**, no imaginary frequencies are observed in the phonon spectra of the five monolayers, confirming their excellent kinetic stability. The AIMD simulation results (**Figure S2**) further validate that these materials can maintain structural integrity under room temperature, demonstrating good thermal stability. Finally, according to the Born-Huang mechanical stability criteria ($C_{11}C_{22} - C_{12}^2 > 0$ and $C_{66} > 0$),[32] the data in **Table S1** indicate that all five materials meet the conditions for mechanical stability. Collectively, these results confirm that the five 2D kagome materials designed in this work can stably exist as free-standing 2D structures, effectively addressing the core issue of their synthesis feasibility from a theoretical perspective and laying the foundation for subsequent experimental synthesis and performance exploration.

**Table 1.** Calculated lattice constant *a*/*b* (Å), atomic thickness *h* (Å), bond length *l* (Å), Young's modulus *Y* (N/s) and Poisson's ration *v* of the $B_{18}S_8$, $B_{18}S_8H_2$ and $B_{18}S_6X_2$ monolayers. The corresponding results of $B_{18}S_6$ are given for comparison.[26]

| Materials | *a*/*b* | *h* | $l_{B-B}$ (in-layer) | $l_{B-B}$ (inter-layer) | $l_{B-S1/S2}$ | $l_{B-X}$ | *Y* | *v* |
|---|---|---|---|---|---|---|---|---|
| $B_{18}S_6$ | 9.14 | 3.190 | 1.70 | 1.73 | 1.88 | -- | 111.63 | 0.24 |
| $B_{18}S_8$ | 9.056 | 3.227 | 1.713/1.763 | 1.748/1.759 | 1.878/1.809 | -- | 60.326 | 0.305 |
| $B_{18}S_8H_2$ | 9.449 | 5.654 | 1.698/1.762 | 1.744/1.761 | 1.818/1.808 | -- | 64.088 | 0.345 |
| $B_{18}S_6Cl_2$ | 9.073 | 3.373 | 1.690/1.799 | 1.709/1.798 | 1.812 | 1.905 | 54.743 | 0.280 |
| $B_{18}S_6Br_2$ | 9.250 | 3.418 | 1.692/1.801 | 1.707/1.798 | 1.812 | 2.052 | 50.262 | 0.281 |
| $B_{18}S_6I_2$ | 9.039 | 3.856 | 1.698/1.797 | 1.708/1.795 | 1.813 | 2.216 | 51.952 | 0.280 |

Based on the elastic constants, we calculated the angle-dependent Young's modulus (*Y*) and Poisson's ratio (*v*) of five kagome monolayers, with the results shown in **Figures 1d** and **1e**. Compared with the parent material $B_{18}S_6$, the Young's modulus of the five newly obtained derivative monolayers is approximately half of that of $B_{18}S_6$. This data clearly indicates that after the introduction of vacancies, the stiffness of the materials decreases significantly, while the flexibility of the materials is enhanced. Specifically, the Young's modulus values of $B_{18}S_8$ and $B_{18}S_8H_2$ are similar and both are higher than that of $B_{18}S_6X_2$. This phenomenon suggests that the strength of the B-S1 bond is stronger than that of the B-X bond. Further comparison reveals that the Young's modulus of $B_{18}S_8$, $B_{18}S_8H_2$, and $B_{18}S_6X_2$ is much lower than that of materials such as graphene (~340 N/m)[33] and $MoS_2$ (~200 N/m)[34], but is relatively close to that of black phosphorus (~40 N/m)[35]. In addition, as shown in **Table 1** regarding the Poisson's ratio, the Poisson's ratio (ranging from 0.208 to 0.345) of the five derivative monolayers is larger than that of the parent material $B_{18}S_6$ (0.24). This change in Poisson's ratio indicates that the introduction of vacancies enhances the deformation capacity of the materials. Considering the characteristic of relatively low Young's modulus, it can be concluded that these five materials are excellent candidates for flexible devices.

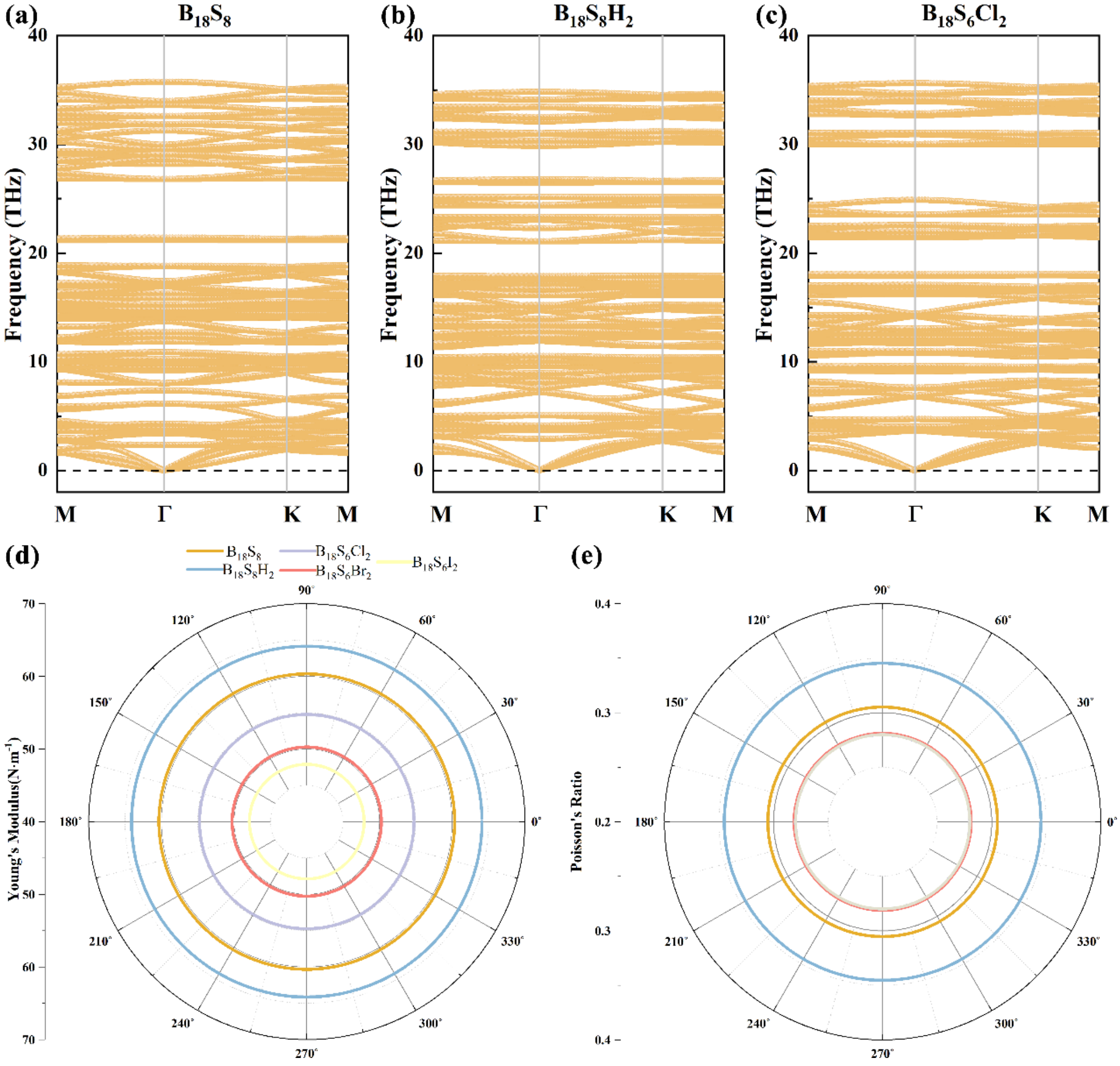

**Figure 2.** Phonon dispersions of (a) $B_{18}S_8$, (b) $B_{18}S_8H_2$ and (c) $B_{18}S_6Cl_2$ monolayers. The calculated angle-dependent in-plane (d) Young's modulus and (e) Poisson's ratio of $B_{18}S_8$, $B_{18}S_8H_2$ and $B_{18}S_6X_2$ monolayers.

The achievements in stability analysis have laid a solid foundation for subsequent studies on physical properties. Based on our well-designed strategy, by introducing “$B_6$” vacancies into $B_{18}S_6$, we successfully obtained $B_{18}S_8$ with typical characteristics of a kagome lattice. Theoretically, its electronic structure should exhibit unique kagome band features, encompassing key characteristics such as Dirac cones (DC), flat bands (FB) and Van Hove singularities (VHS). As shown in **Figure 2a**, calculations based on the GGA-PBE method indicate that $B_{18}S_8$ exhibits metallic properties, in stark contrast to the wide-bandgap semiconductor properties of $B_{18}S_6$ (see **Figure S3** for details). Further analysis reveals that relatively ideal kagome bands are indeed formed in the electronic structure of $B_{18}S_8$. Among them, two quadratic bands and the Dirac cone they form are located in the conduction band region above the Fermi level. Specifically, the Dirac cone at the K point is approximately at 0.64 eV in the conduction band, while the flat band is situated below the Fermi level with a band dispersion of about 0.2 eV. Although the Dirac cone in $B_{18}S_8$ is not at the Fermi level, it is undeniable that the introduction of “$B_6$” vacancies into $B_{18}S_6$ has enabled the system to acquire the band characteristics of a kagome lattice.

Upon delving deeper into the evolution of the electronic structure, we discovered that the parent material $B_{18}S_6$ harbors hidden kagome bands within an energy range of -1.2 to 0 eV (see **Figure S3**). However, due to the influence of orbital coupling, these kagome bands are not perfect. When "B6" cation vacancies are introduced, hole-type vacancies are introduced into the system, causing the Fermi level to shift towards the valence band. This shift results in the upward movement of the kagome bands in $B_{18}S_6$, with some of them entering the conduction band region. Simultaneously, the introduction of "$B_6$" vacancies further accentuates the kagome characteristics of the system, making the kagome band features closer to the ideal state and ultimately forming the electronic structure of $B_{18}S_8$. To precisely move the Dirac cone in $B_{18}S_8$ to the Fermi level, we adopted an effective strategy: hydrogen passivation of the 3-coordinated S1 atoms in $B_{18}S_8$, thereby synthesizing $B_{18}S_8H_2$. **Figure 2b** displays the band structure of $B_{18}S_8H_2$. As expected, compared to $B_{18}S_8$, the position of its kagome bands is more perfect. Specifically, firstly, the Dirac cone is precisely located at the Fermi level, a characteristic similar to that of graphene, providing extremely favorable conditions for subsequent regulation in device applications. Secondly, it possesses "isolated" kagome bands, with no interference from other "impurity" bands in the vicinity in the band structure, facilitating effective experimental observation and practical application. The successful acquisition of $B_{18}S_8H_2$ provides a highly valuable example for further expanding derivative materials. Building on this, we replaced the "-SH" groups in $B_{18}S_8H_2$ with halogen atoms and successfully obtained the $B_{18}S_6X_2$ (X = Cl, Br, I) series of materials while ensuring the balance of valence electrons in the system. **Figure 2c** presents the band results of $B_{18}S_6Cl_2$, while the band structures of $B_{18}S_6Br_2$ and $B_{18}S_6I_2$ are shown in **Figure S4**. As anticipated, the kagome band characteristics of $B_{18}S_6X_2$ near the Fermi level are almost identical to those of $B_{18}S_8H_2$. The difference lies in that $B_{18}S_6X_2$ also exhibits relatively ideal kagome bands in the conduction band region. For example, the kagome bands of $B_{18}S_6Cl_2$ are located within an energy range of 1.4 to 2.5 eV (see **Figure 3c**).

Subsequently, we thoroughly examined the impact of the spin-orbit coupling (SOC) effect on the electronic structures of the predicted materials. At the GGA-PBE calculation level, the SOC effect opens a bandgap of approximately 0.07 eV near the Γ point in the $B_{18}S_8$ monolayer. For $B_{18}S_8H_2$ and $B_{18}S_6X_2$, the SOC effect has almost no significant impact on its electronic structure (see **Figures 3b** and **S4** for details). Considering the limitations of the GGA-PBE method in predicting band structures, we further recalculated the band structures of the studied materials using the hybrid functional HSE06, taking into account the influence of the SOC effect. As shown in **Figures 3d-3f** and **Figure S4**, without considering the SOC effect, the band characteristics of $B_{18}S_8$ are basically consistent with the GGA-PBE calculation results. However, when the SOC effect is considered, its band structure undergoes splitting. Specifically, a bandgap of approximately 0.052 eV is opened at the Dirac cone at the K point, and a tiny bandgap of about 0.085 eV is also opened near the Γ point. For $B_{18}S_8H_2$, the band results obtained through hybrid functional calculations show that the SOC effect has almost no impact on the kagome bands at the Fermi level of the system. Compared to the GGA-PBE results, the main difference is that the Dirac cone at the K point opens a tiny bandgap (about 0.025 eV). For the $B_{18}S_6X_2$ monolayers, the influence of the hybrid functional and the SOC effect on the kagome bands near the Fermi level is similar to that of $B_{18}S_8H_2$. Among them, the bandgaps opened by the SOC effect at the K point are 0.051 eV ($B_{18}S_6Cl_2$), 0.055 eV ($B_{18}S_6Br_2$), and 0.051 eV ($B_{18}S_6I_2$), respectively. These values are much larger than the bandgap opened at the Dirac cone of graphene under the SOC effect (only 1 meV). From the perspective of the practical requirements of device applications, a larger bandgap helps achieve a higher on-off ratio, thereby improving device performance. Based on the above research results, the $B_{18}S_8H_2$ and $B_{18}S_6X_2$ series of materials exhibit enormous potential and are expected to be candidate materials for high-performance low-dimensional devices.

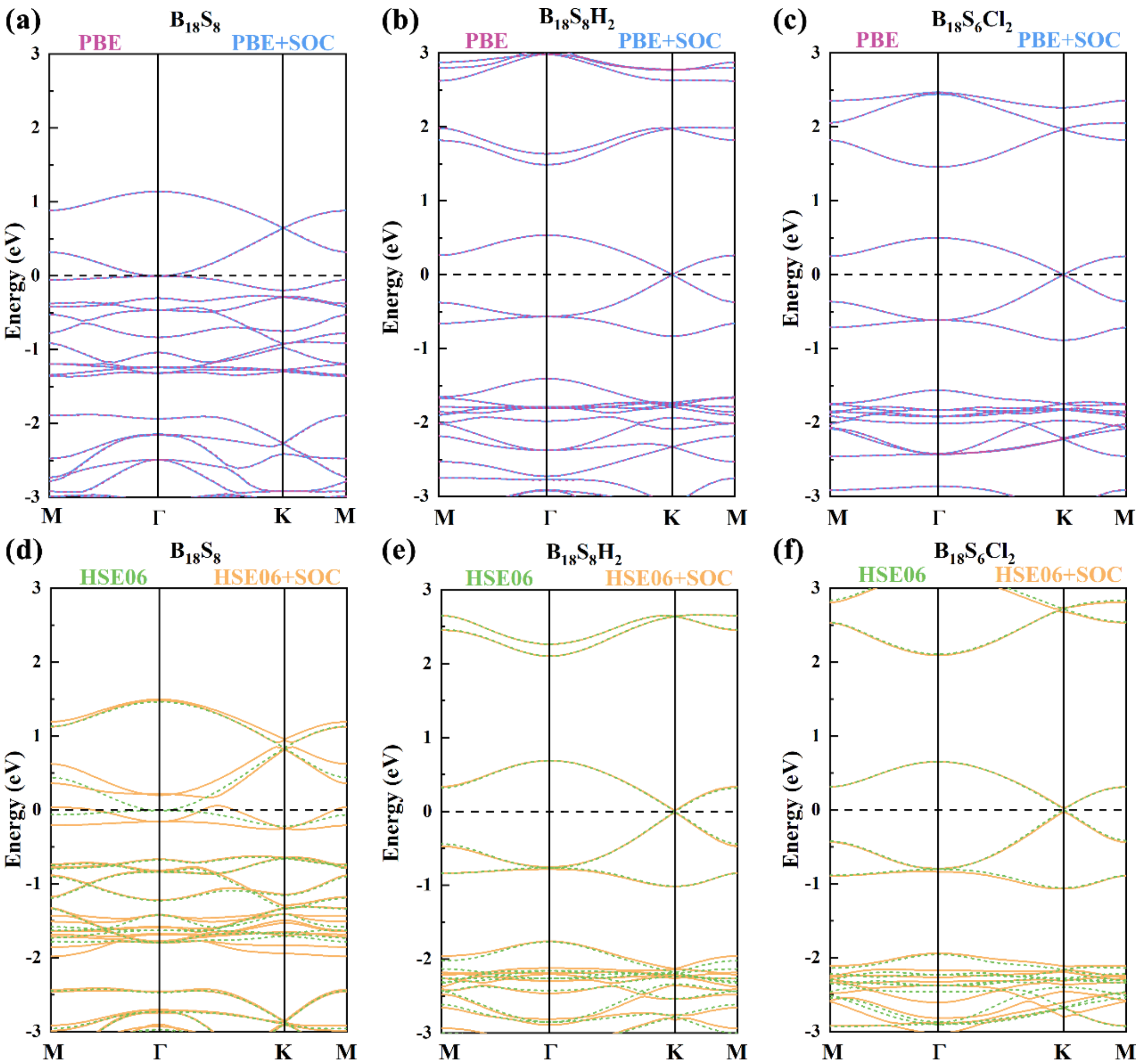

**Figure 3.** Calculated band structures of (a, d) $B_{18}S_8$, (b, e) $B_{18}S_8H_2$ and (c, f) $B_{18}S_6Cl_2$ monolayers based on GGA-PBE/HSE06 computational methods with/without SOC effect considered.

Next, we will focus on analyzing the specific contributions of each atomic orbital to the band structure and density of states (DOS) in the three types of materials: $B_{18}S_8$, $B_{18}S_8H_2$, and $B_{18}S_6X_2$. The relevant results are presented in **Figure 4**. From **Figure 4a**, it can be clearly observed that in the $B_{18}S_8$ monolayer, the main contributions to the kagome bands come from the "$B_6$" atomic clusters and the S2 atoms in a 2-coordinate state. Specifically, these contributions primarily originate from the in-plane $p_x$ and $p_y$ orbitals, as well as the out-of-plane $p_z$ orbital, of both B and S2 atoms. This characteristic significantly differs from that of graphene, where the Dirac cone contribution solely comes from the out-of-plane C-$p_z$ orbital. Through the analysis of the electronic DOS, it is evident that the presence of flat bands near the Fermi level leads to highly localized DOS features, while the DOS exhibits a linear increase near the Dirac cone. Additionally, the Van Hove singularity located at the M point in the kagome bands induces two prominent peaks in the DOS.

For the $B_{18}S_8H_2$ material (see **Figure 4b**), the orbital contributions to the kagome bands are essentially consistent with those in $B_{18}S_8$, primarily stemming from the $p_x$, $p_y$, and $p_z$ orbitals of B and S2 atoms. Furthermore, the corresponding DOS plot effectively illustrates the DOS distribution characteristics associated with the Dirac cone, Van Hove singularity, and flat band. In the $B_{18}S_6X_2$ monolayer, as shown in **Figures 4c**, **S5** and **S6**, the orbital contributions to the kagome bands and the corresponding electronic DOS features are similar to those in $B_{18}S_8$ and $B_{18}S_8H_2$. However, there is a notable difference between $B_{18}S_6X_2$ and the previous two materials: in the conduction band of $B_{18}S_6X_2$ (for example, for $B_{18}S_6Cl_2$, the conduction band ranges from 1.5 to 2.5 eV), there exists a kagome band jointly contributed by B and X atoms. This kagome band is mainly contributed by the $s$, $p_x$, $p_y$, and $p_z$ orbitals of B and X atoms, and the corresponding DOS plot also clearly reflects the DOS characteristics associated with the Dirac cone, Van Hove singularity, and flat band. Nevertheless, compared to the kagome bands at the Fermi level, those located at higher energy levels in the conduction band pose significant challenges for practical observation and application. Therefore, we will not delve further into them in subsequent research discussions.

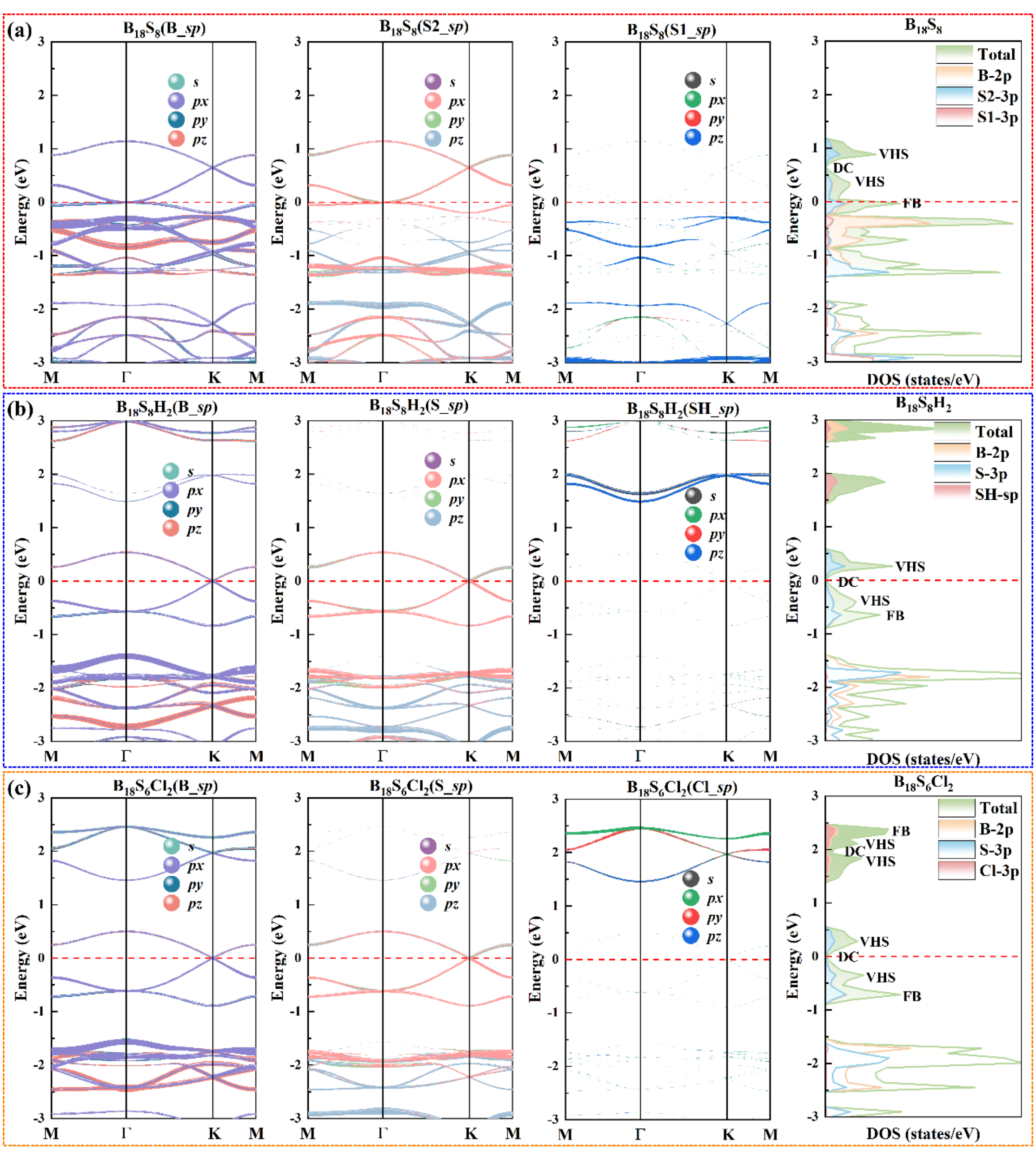

**Figure 4.** Projected band structures density of states of (a) $B_{18}S_8$, (b) $B_{18}S_8H_2$ and (c) $B_{18}S_6Cl_2$ monolayers at GGA-PBE level without SOC effect considered.

Last but not least, based on the computational results obtained via GGA-PBE method, we plotted the 3D band diagrams corresponding to the kagome bands of five novel materials and calculated the Fermi velocities near their Dirac cones. The relevant results are shown in **Figure 5**. From the 3D band diagrams, it can be clearly observed that the bands near the Dirac cones of these five materials all exhibit good linear characteristics, forming nearly ideal Dirac cone structures. This indicates that massless Dirac fermions exist at the Dirac cones in these systems.

The calculated Fermi velocity results reveal that $B_{18}S_8$ has the lowest Fermi velocity, with values of $2.716\times10^5$ m/s and $2.690\times10^5$ m/s in the Γ-K and K-M directions, respectively, as shown in **Figure 5f** and **Table S2**. In contrast, the Dirac cone in $B_{18}S_8H_2$ exhibits higher linearity compared to that in $B_{18}S_8$, resulting in higher Fermi velocities of $2.947\times10^5$ m/s and $2.930\times10^5$ m/s in the Γ-K and K-M directions, respectively. For the $B_{18}S_6X_2$ monolayers, their Fermi velocities show a periodic increasing trend. Among them, $B_{18}S_6Cl_2$ has the lowest Fermi velocity, with values of $2.873\times10^5$ m/s and $2.840\times10^5$ m/s in the Γ-K and K-M directions, respectively, while $B_{18}S_6I_2$ has the highest Fermi velocity, reaching $3.067\times10^5$ m/s and $3.046\times10^5$ m/s in the Γ-K and K-M directions, respectively. Notably, the Fermi velocities of $B_{18}S_8$, $B_{18}S_8H_2$, and $B_{18}S_6X_2$ are comparable to that of graphene ($8.22\times10^5$ m/s) [36], fully demonstrating their excellent carrier transport properties. However, graphene's zero-bandgap characteristic makes it difficult to achieve a high on-off ratio, which to some extent limits its application in high-performance electronic devices. In contrast, $B_{18}S_8H_2$ and $B_{18}S_6X_2$ can open bandgaps of 25-55 meV under the influence of SOC effects. This characteristic not only ensures high carrier mobility in the materials but also lays the foundation for designing high on-off ratio devices, indicating their broad and excellent application potential in the field of future electronic devices.

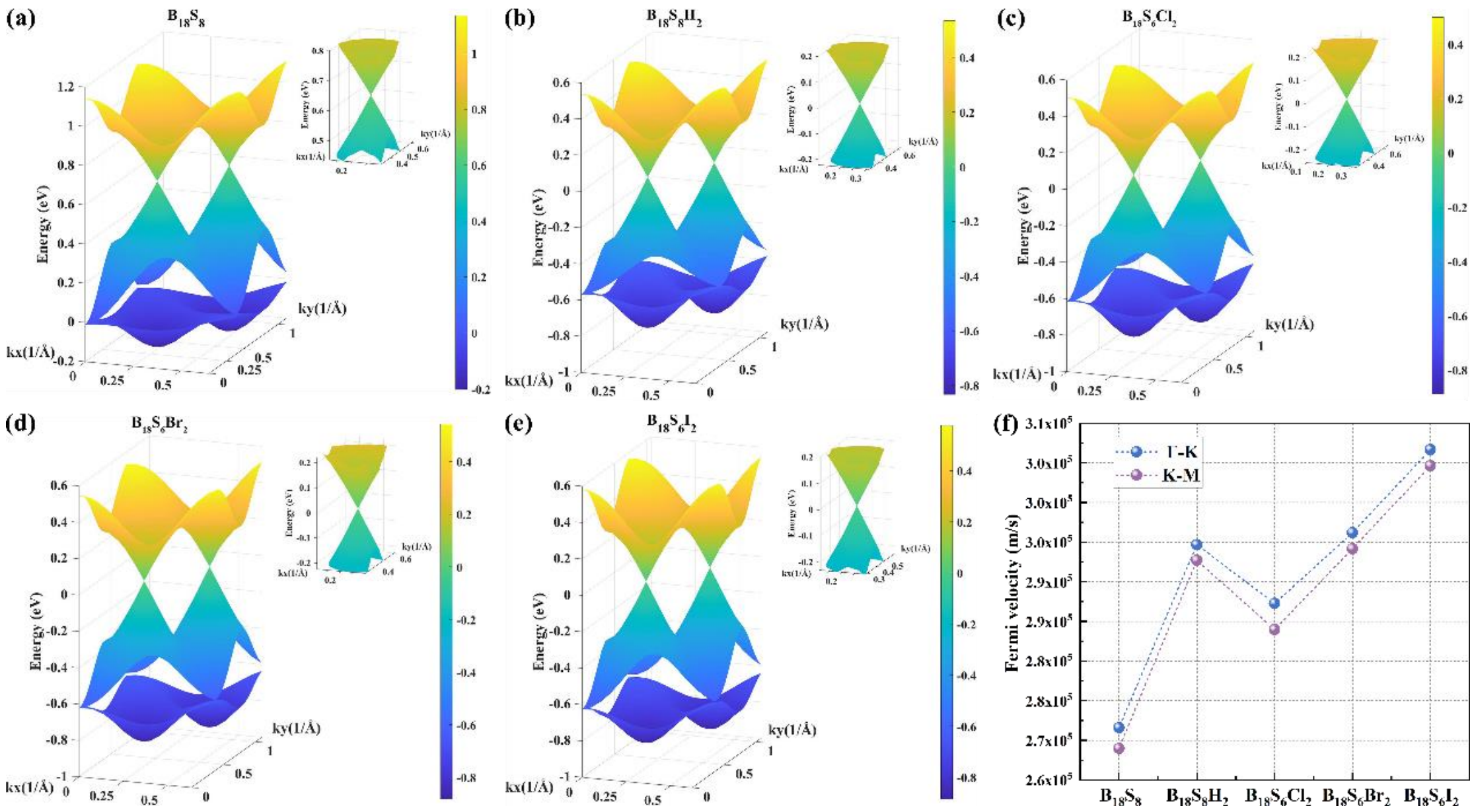

**Figure 5.** 3D band structures of (a) $B_{18}S_8$, (b) $B_{18}S_8H_2$, (c) $B_{18}S_6Cl_2$, (d) $B_{18}S_6Br_2$ and (e) $B_{18}S_6I_2$ monolayers based on GGA-PBE calculations. (f) The corresponding Fermi velocity of $B_{18}S_8$, $B_{18}S_8H_2$, and $B_{18}S_6X_2$ monolayers around the Dirac cones along Γ-K and K-M directions.

## 4. Conclusion

In summary, we harnessed the "1+3" design strategy proposed in our earlier work, along with surface passivation and charge balance techniques, to successfully devise a novel family of 2D kagome materials, namely $B_{18}S_8$, $B_{18}S_8H_2$, and $B_{18}S_6X_2$ (X = Cl, Br, I). The stability of these materials is determined by phonon spectra, ab initio molecular dynamics simulations, and elastic constants. These five materials exhibit relatively low

Young's moduli (ranging from 50 to 64 N/m) and relatively high Poisson's ratios (ranging from 0.28 to 0.34), making them excellent flexible materials. Electronic structure analysis unveiled that while $B_{18}S_8$ showcases remarkable kagome band characteristics, its Dirac cone is situated roughly 1 eV above the Fermi level. Nevertheless, through surface hydrogen passivation, the Dirac cone was able to effectively shift to the Fermi level. Further exploration revealed that substituting surface sulfur atoms with halogen atoms could similarly reposition the Dirac cone at the Fermi level as well. For these five materials, the Fermi velocities in the vicinity of the Dirac cone soar to an impressive range of 2.69 to $3.07\times10^5$ m/s. Moreover, spin-orbit coupling can induce a bandgap of approximately 25 to 55 meV at the Dirac cone. Our research not only serves as an exemplary model for the design of 2D boron-based kagome materials but also fully underscores the vast potential these materials hold in the realm of electronics.

**Acknowledgements**

This work was supported by the National Undergraduate Innovation and Entrepreneurship Training Program (Grant No. 202510497080).